\documentclass[prd,showkeys,floatfix,twocolumn,amsmath,amssymb,floatfix]{revtex4}
\usepackage{graphicx}
\usepackage{multirow}
\usepackage{subfigure}
\usepackage[colorlinks,citecolor=blue,anchorcolor=red,menucolor=red,linkcolor=red,filecolor=red,runcolor=red,urlcolor=blue,frenchlinks=red]{hyperref}
\usepackage{enumitem}
\usepackage{slashed}
\usepackage[normalem]{ulem}

\makeatletter
\renewcommand{\@thesubfigure}{\hskip\subfiglabelskip}
\makeatother
\allowdisplaybreaks[3]

\begin{document}
\title{Investigation on the $\Omega(2012)$ from QCD sum rules}
\author{Niu Su$^{1,2}$}
\email{suniu@rcnp.osaka-u.ac.jp}
\author{Hua-Xing Chen$^2$}
\email{hxchen@seu.edu.cn}
\author{Philipp Gubler$^3$}
\email{philipp.gubler1@gmail.com}
\author{Atsushi Hosaka$^{1,3}$}
\email{hosaka@rcnp.osaka-u.ac.jp}

\affiliation{$^1$Research Center for Nuclear Physics (RCNP), Osaka University, Ibaraki 567-0047, Japan\\
$^2$School of Physics, Southeast University, Nanjing 210094, China\\
$^3$Advanced Science Research Center, Japan Atomic Energy Agency (JAEA), Tokai 319-1195, Japan
}
\begin{abstract}
We study the recently observed $\Omega(2012)$ baryon in QCD sum rules. We construct the $P$-wave $\Omega$ baryon currents with a covariant derivative, and perform spin projection to obtain the currents with total spin 1/2 and 3/2. We then apply the parity-projected QCD sum rules to separate the contributions of the positive and negative parity states. We extract the masses of $J^P = 1/2^-$ and $3/2^-$ states to be $M_{1/2^-} = 2.07^{+0.07}_{-0.07}{\rm~GeV}$ and $M_{3/2^-} = 2.05^{+0.09}_{-0.10}{\rm~GeV}$. Both results are in good agreement with the experimental result. 
Therefore, it is likely that the $\Omega(2012)$ is a negative parity state, which is interpreted as a $P$-wave excited state in the quark model.  
However, its spin is not determined in the present analysis, which can be done by detailed study on its decay properties.  

\end{abstract}

\maketitle

\section{Introduction}
\label{sec:intro}

Significant progress has been recently achieved in baryon spectroscopy, with an increasing number of new baryon candidates, especially for those containing heavy quarks~\cite{LHCb:2012kxf,CDF:2013pvu,LHCb:2017uwr,Belle:2017ext,LHCb:2020iby,CMS:2021rvl,LHCb:2021ptx,LHCb:2023zpu}. However, even our understanding of the properties of light baryons is still incomplete, especially for those containing strange quarks. With the observation of more $\Xi$ and $\Omega$ baryons~\cite{BESIII:2015dvj,Belle:2018mqs,Belle:2018lws,Belle:2019zco,BESIII:2019cuv,LHCb:2020jpq,Belle:2021gtf,Belle:2022mrg} together with a future hadron facility plan at, for instance J-PARC~\cite{Aoki:2021cqa}, it is now the good time to study light baryons containing strange quarks in more detail. Furthermore, by accumulating our knowledge of strange hadrons, we can improve our understanding of the nonperturbative properties of QCD.

In 2018, the exited $\Omega$ baryon, $\Omega(2012)$, was observed for the first time in  $\Xi^0K^-$ and $\Xi^-K^0_s$ invariant mass distributions in $\Upsilon(1S), \Upsilon(2S)$, and $\Upsilon(3S)$ decays by the Belle experiment~\cite{Belle:2018mqs}.
The experimental evidence has been further strengthened by the 
$\Omega_c \to \pi^+\Omega(2012)^- \to \pi^+(\bar K \Xi)^-$ decay~\cite{Belle:2021gtf}. 
The latest data for its mass and decay width are~\cite{Belle:2022mrg}:
\begin{eqnarray}
M &=& 2012.5 \pm 0.7 \pm 0.5 {\rm~MeV} \, ,
\\   \nonumber
\Gamma &=& 6.4^{+2.5}_{-2.0}  {\rm~MeV} \, .
\end{eqnarray}
Furthermore, due to the relatively narrow width it has been argued that the spin-parity of $\Omega(2012)$
is likely to be $J^P = 3/2^-$. 

The conventional quark model may naively explain the $\Omega(2012)$ to be a negative parity state
as the first $P$-wave excitation of the ground-state $\Omega$ baryon with three strange quarks~\cite{Aliev:2018syi,Aliev:2018yjo,Polyakov:2018mow,Wang:2018hmi,Xiao:2018pwe,Liu:2019wdr,Arifi:2022ntc,Menapara:2021vug,Wang:2022zja,Zhong:2022cjx}. 
One important feature of this quark model picture is that there should be spin-orbit partners of both $J^P = 1/2^-$ and $3/2^-$. 
The spin $1/2^-$ state is expected to be rather wide, which could be the reason that only one narrow state of $3/2^-$ has been observed so far. 

In contrast, because of the fact that the mass of $\Omega(2012)$ is close to the
$\bar K$ and $\Xi^*(1530)$ threshold, a molecular picture of these particles
has been proposed and extensively discussed in Refs.~\cite{Lin:2018nqd,Valderrama:2018bmv,Pavao:2018xub,Huang:2018wth,Gutsche:2019eoh,Ikeno:2020vqv,Zeng:2020och,Lu:2020ste,Liu:2020yen,Ikeno:2022jpe,Hu:2022pae}.
This picture is attractive in relation with the recent discussions of exotic hadrons appearing near thresholds~\cite{Chen:2016qju,Guo:2017jvc,Ali:2017jda,Olsen:2017bmm,Liu:2019zoy,Guo:2019twa,Chen:2022asf,Liu:2024uxn}. 
Unlike the quark model picture, however, the spin $1/2^-$ state is not easily generated in the molecular picture.   
The molecular structure of $\Xi^*(1530) \bar K$ may furthermore lead to a large contribution to the three-body decay of
$\Omega(2012) \to \Xi^*(1530) \bar K \to \Xi \pi \bar K$~\cite{Lin:2018nqd,Valderrama:2018bmv,Pavao:2018xub,Huang:2018wth}. 
In experiments, however, it was first reported that such a three-body decay was not observed~\cite{Belle:2019zco}, but later the measurement was revisited and the possibility of the three-body decay was discussed in Ref.~\cite{Belle:2022mrg}. 

In this situation, a hybrid picture of three-quark and molecular structures was proposed~\cite{Lu:2022puv}. In that work, 
the three-quark component (referred to as the bare component) was reported to contribute only about 29 \% to the physical 
$\Omega(2012)$ state. The relatively small rate is however due to the mixture of a non-interacting $\bar K \Xi^*$ one-loop component in the physical state and does not necessarily mean molecular dominance.  
After all, at this moment it would be fair to say that the structure of $\Omega(2012)$ is not yet well understood, and this has motivated us to study further properties of this state in yet another theoretical approach based on QCD sum rules. 

The starting point of QCD sum rules is to construct an interpolating current corresponding to the state of interest. 
There are former works for the $\Omega(2012)$ using a local three-quark current without derivative~\cite{Aliev:2018syi,Aliev:2018yjo}, but in this paper we employ a 
current
with a covariant derivative, assuming that the parity of $\Omega(2012)$ is negative.
In this way we expect that the current couples more strongly to physical states better than the one without derivative. 
We then perform a systematic study for the four possible spin-parity states $J^P = 1/2^{\pm}$ and $3/2^{\pm}$,
by properly applying spin and parity projections. As a result, we extract the masses of the $1/2^-$ and $3/2^-$ states 
with values around the observed mass of the $\Omega(2012)$, namely 
\begin{eqnarray}
M_{1/2^-} &=& 2.07^{+0.07}_{-0.07}{\rm~GeV} \, ,
\end{eqnarray}
and
\begin{eqnarray}
M_{3/2^-} &=& 2.05^{+0.09}_{-0.10}{\rm~GeV} \, ,
\end{eqnarray}
while the masses for the positive parity states appear in a much higher region around 3 GeV with larger uncertainties. 
Our present results imply that the $\Omega(2012)$ is dominated  by a three-quark state. However, to reach the final conclusion we still need further studies: (1) compute decay width, and (2) use the five-quark current corresponding to the molecular state $\Xi^*(1530) \bar K$. We relegate these two issues to future works. 

This paper is organized as follows. In Sec.~\ref{sec:current} we explicitly construct the $P$-wave currents for the $\Omega$ baryon with a covariant derivative. Then we perform a parity-projected QCD sum rule analysis for them in Sec.~\ref{sec:sumrule}. In Sec.~\ref{sec:numerical}, we provide and discuss the results of the numerical analysis. The obtained findings are summarized in Sec.~\ref{sec:summary}.

\section{$P$-wave $\Omega$ baryon currents}
\label{sec:current}

Let us first construct the currents for the $P$-wave
$\Omega$ baryon using three strange quark fields $s_a(x)$ which have a (not explicitly written) Dirac spinor structure with $a = 1\cdots3$ being color indices.  
The logic for constructing the current goes in a way similar to the Ioffe argument~\cite{Ioffe:1981kw} supplemented by a non-vanishing diquark structure.
In the presence of a derivative, it can be shown that three types of
diquarks remain non-vanishing:
$\epsilon^{abc} s_a^{T} C \gamma_5 {\overset{\leftrightarrow}{D}}_\mu s_b $, $\epsilon^{abc} s_a^{T} C {\overset{\leftrightarrow}{D}}_\mu s_b$ and 
$\epsilon^{abc} s_a^{T} C \gamma_\mu \gamma_5 {\overset{\leftrightarrow}{D}}_\mu s_b$, which are the scalar, vector, and axial-vector types, respectively.  
Here $D_\mu = \partial_\mu + i g_s A_\mu$ is the covariant derivative with the gluon field $A_\mu$.
Furthermore, for baryon fields of structure $\epsilon^{abc} (s_a^{T} \Gamma_1 s_b)\Gamma_2 s_c$, Fierz rearrangement  shows that only two of them are independent, for which we may choose 
vector and axial-vector types.  
As discussed in Ref.~\cite{Ioffe:1981kw}, one of the linear combinations has the advantageous property of surviving in the non-relativistic limit 
and has the preferable chiral structure of contributing to the important 
terms in the operator product expansion of the two-point function, that are proportional to the quark condensate, 
for instance $\langle \bar s_L s_R\rangle$.

In this work we employ, among the above shown three diquarks, the following $ss$-diquark
\begin{equation}
\epsilon^{abc} [s_a^T C \gamma_5 {\overset{\leftrightarrow}{D}}_\mu s_b] = - 2 \epsilon^{abc} [(D_\mu s_a^T) C \gamma_5 s_b] \, 
\label{eq_diquark_Our_Choice}, 
\end{equation}
This diquark has a suitable internal $P$-wave structure.  In fact, $s_a^T C \gamma_5 s_b$ has quantum number $J^P=0^+$ which is $S$-wave, with total spin $s_{12} = 0$. After applying a derivative, $s_a^T C \gamma_5{\overset{\leftrightarrow}{D}}_\mu s_b$ becomes a pure $P$-wave diaqurk with orbital angular momentum $l_{\rho} = 1$, and total angular momentum $j_{12} = l_\rho + s_{12} = 1$. In contrast, the other two candidates, the diquarks $s_a^T C s_b$ and $s_a^T C  \gamma_\mu \gamma_5 s_b$, have quantum numbers $J^P=0^-$ and $1^-$, respectively, which have  already $P$-wave nature. When a derivative is applied, the diquarks, $s_a^{T} C {\overset{\leftrightarrow}{D}}_\mu s_b$ and $s_a^{T} C \gamma_\mu \gamma_5 {\overset{\leftrightarrow}{D}}_\mu s_b$, obtain complicated structure unlike the $P$-wave field of our choice (\ref{eq_diquark_Our_Choice}).  
The use of these diquarks for the baryon current is expected to be less reliable. 
In fact, we have explicitly verified that the sum rule using the second diquark for the baryon currents does not result in good signals for negative parity baryons around 2 GeV, but rather to signals  at around 3 GeV.  Because the baryon current with the third diquark is not independent, and hence can be expressed by the other two baryon currents, it does not alter our conclusion.  
Therefore, we employ the diquark $s_a^{T} C \gamma_5 {\overset{\leftrightarrow}{D}}_\mu s_b$ to construct the baryon currents. We note here that for the ground octet and decuplet baryons, it can be advantageous to consider the combinations of the currents without derivatives~\cite{Leinweber:1995ie,Leinweber:1994nm,Ioffe:1982ce,Chung:1981cc,Espriu:1983hu}.

Combining the diquark with the third quark field of spin $1/2$,
one can write the currents for the $P$-wave $\Omega$ baryon with the total angular momentum $J_{tot} = 1/2,\ 3/2$ correspondingly as following
\begin{eqnarray}
J &=& -2\epsilon^{abc} ~ [(D^\mu s_a^T) C \gamma_5 s_b]
\label{def:current1}~ \gamma_\mu s_c \, ,
\\ 
J_\mu &=& -2\epsilon^{abc} ~ [(D^\nu s_a^T) C \gamma_5 s_b] \label{def:current2}~ (g_{\mu\nu} - 
{1\over4}\gamma_\mu\gamma_\nu) s_c \, .
\end{eqnarray}    
Because these two currents have the above-mentioned preferable properties, we will employ them in the present study.

\section{QCD sum rule analyses}
\label{sec:sumrule}

In this section, we briefly explain the method of QCD sum rules~\cite{Shifman:1978bx,Shifman:1978by,Reinders:1984sr,Narison:2002woh,Nielsen:2009uh,Gubler:2018ctz} by using the current listed in Eq.~(\ref{def:current2}). 
Let us start with the current matrix element.
As an example, the current $J_\mu$, whose spin and parity are $J^P = 3/2^-$ can couple to the physical state $|\Omega; 3/2^-\rangle$ with the relevant matrix element written as 
\begin{eqnarray}
\langle 0 | J_\mu | \Omega; 3/2^- \rangle = f_{-} u_\mu(q)\, ,
\label{coupling1}
\end{eqnarray}
where 
$f_-$ is a coupling  constant and $u_\mu(q)$ the Rarita-Schwinger vector-spinor. The current $J_\mu$ can also couple to a positive parity  state $|\Omega; 3/2^+ \rangle$ 
whose matrix element is given as 
\begin{eqnarray}
\langle 0 | J_\mu | \Omega;3/2^+ \rangle = f_{+} \gamma_5 u_\mu(q)\, .
\label{coupling2}
\end{eqnarray}
We note that the partner current $\gamma_5 J_\mu$ can also couple to the same physical states, and give the same results as $J_\mu$.    

We next study the correlation function with the following Lorentz structure:
\begin{eqnarray}
\Pi_{\mu\nu}(q^2) &=& i \int d^4x e^{iqx} \langle 0 | {\bf T}[J_{\mu}(x) J_\nu^\dagger(0)] | 0 \rangle \label{twoponit}
\\ \nonumber
&=&(\frac{q_\mu q_\nu}{q^2}-g_{\mu\nu})\Pi(q^2)+\cdots \, .
\end{eqnarray} 
The $\cdots$ represents the contributions from states with spin 1/2 which we neglect here. $\Pi(q^2)$ can also be expressed as a dispersion relation, 
\begin{eqnarray}
\Pi(q^2) = \int^\infty_{s_<}\frac{\rho(s)}{s-q^2-i\varepsilon}ds \, ,
\end{eqnarray}
where $\rho(s) \equiv {\rm Im} \Pi(s)/\pi$ is the spectral density, and $s_< = 9 m_s^2$ is the lower threshold of the spectral function computed by the OPE.

At the hadron level, we obtain the spectral density by inserting the complete set of intermediate hadronic states 
\begin{eqnarray}
\rho^{\rm phen}(s)  &\equiv& \sum_n\delta(s-M^2_n) \langle 0| J_{\mu} | n\rangle \langle n| J_{\nu}^{\dagger} |0 \rangle \label{eq:rho}
\\ \nonumber 
&=&f_-^2 (\slashed{q}+M_-) \delta(s-M_-^2) 
\\ \nonumber
&&+ f_+^2 (\slashed{q}-M_+) \delta(s-M_+^2)
\\ \nonumber
&&+ \theta(s-s_0)\rho^{\rm cont}(s) \, ,
\end{eqnarray}
in which we consider two poles for both $|\Omega; 3/2^-\rangle$ and $|\Omega; 3/2^+\rangle$ as well as the continuum contribution. $M_{\mp}$ are the physical masses of negative and positive parity states, respectively.
Here $s_0$ is the threshold value beyond which ($s > s_0$) the spectral density is approximated by that of OPE (see also the discussion around Eq.~(\ref{sumrule}).  
Based on Eqs.~(\ref{coupling1})-(\ref{eq:rho}), the correlation function can be given as
\begin{eqnarray}
\nonumber \Pi^{\rm phen}(q^2)&\!\!=\!\!&f_-^2 \frac{\slashed{q}+M_-}{M_-^2-q^2 - i \epsilon}+ f_+^2 \frac{\slashed{q}-M_+ }{M_+^2-q^2 - i \epsilon}  
\\ &\!\!=\!\!& \Pi_1^{\rm phen}(q^2)\slashed{q}+\Pi_0^{\rm phen}(q^2)\, ,
\end{eqnarray}
By introducing the spectral densities $\rho^{\rm phen}_{0,1}$ for $\Pi^{\rm phen}_{0,1}$, we can write down the following relations
\begin{eqnarray}
\rho^{\rm phen}_1(s)\!\!&=&\!\!f_-^2\delta(s-M^2_-)+ f_+^2\delta(s-M^2_+)  \label{phen1}\, ,
\\\rho^{\rm phen}_0(s)\!\!&=&\!\!f_-^2M_-\delta(s-M^2_-)- f_+^2M_+\delta(s-M^2_+) \label{phen0}\, , 
\end{eqnarray}
from which we can extract the spectral densities for negative and positive parity states as 
\begin{eqnarray}
\rho^{\rm phen}_{\mp}(s) = \sqrt{s} \rho^{\rm phen}_1(s)\pm \rho^{\rm phen}_0(s) .
\end{eqnarray}

\begin{figure}[hbtp]
\begin{center}
\scalebox{0.12}{\includegraphics{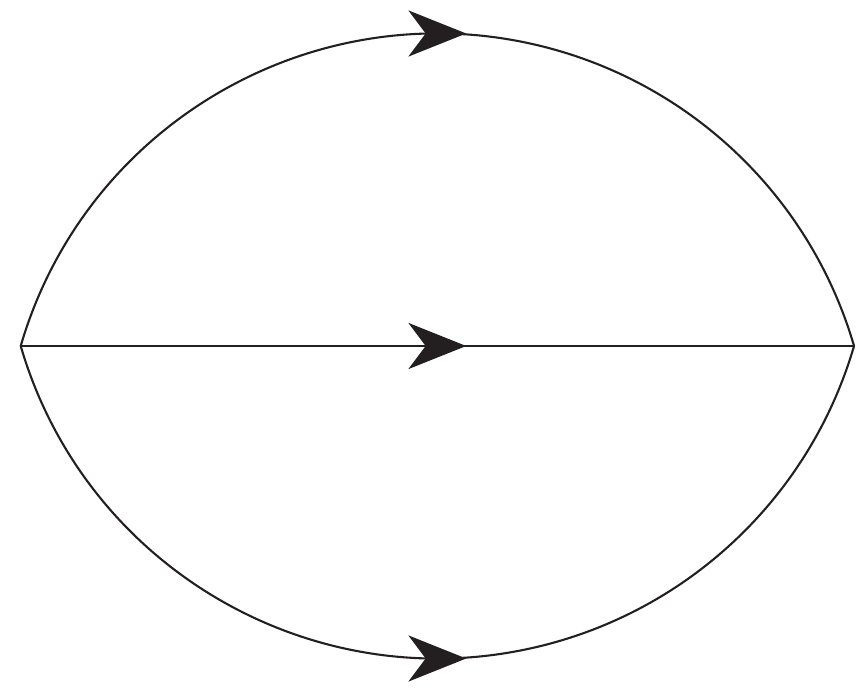}}
\\[2mm]
\scalebox{0.12}{\includegraphics{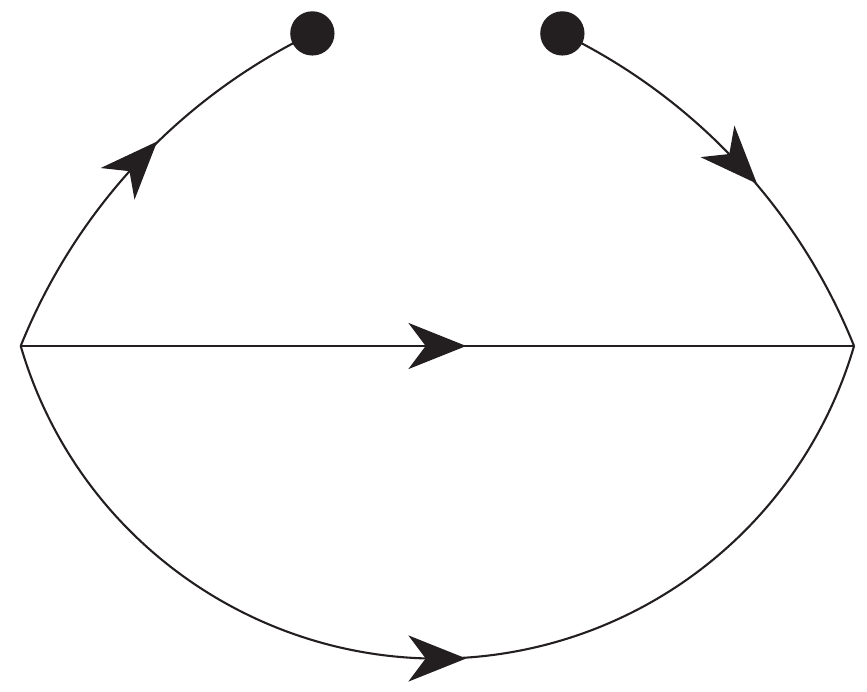}}~~
\scalebox{0.12}{\includegraphics{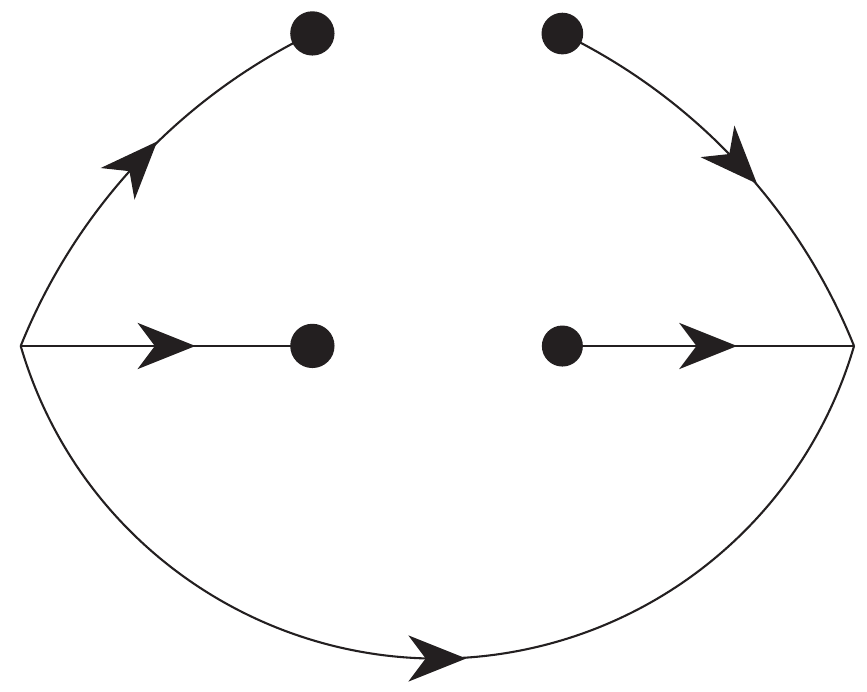}}~~
\\[2mm]
\scalebox{0.12}{\includegraphics{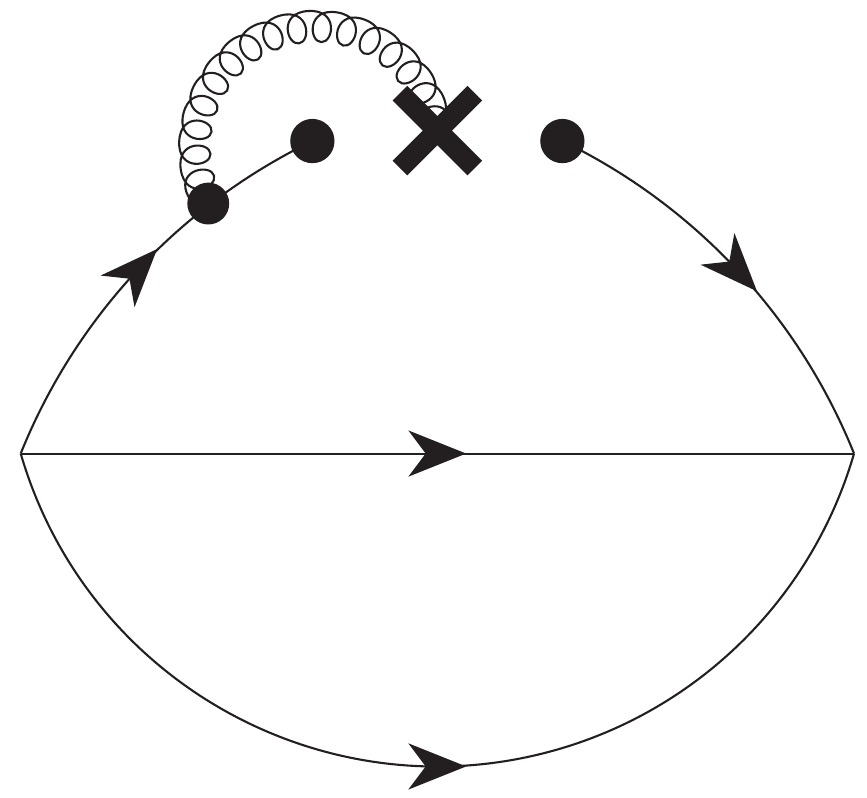}}~~
\scalebox{0.12}{\includegraphics{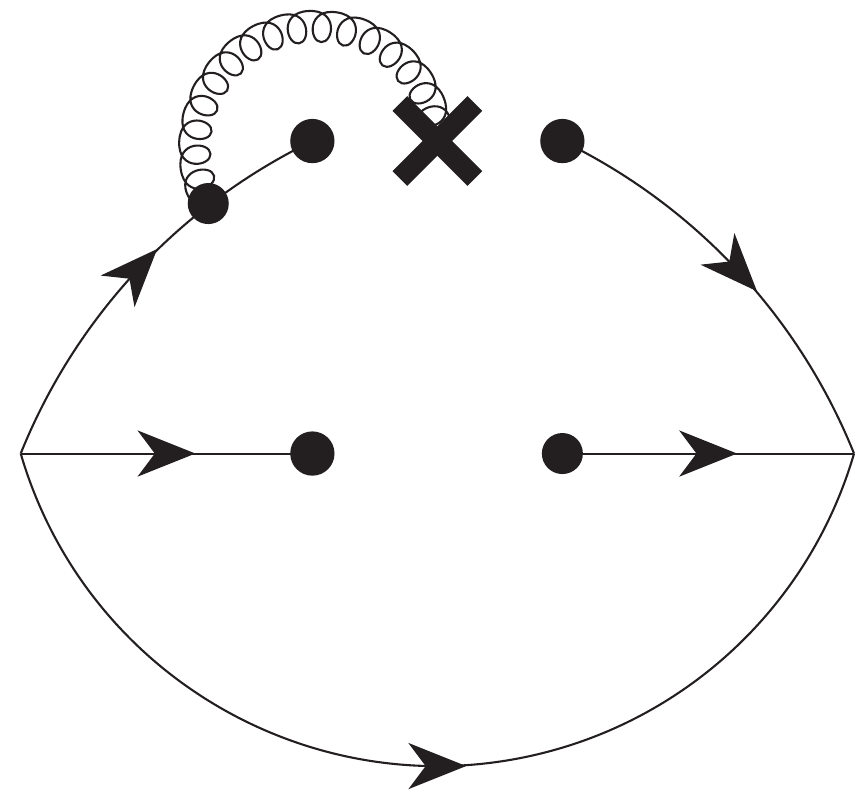}}~~
\scalebox{0.12}{\includegraphics{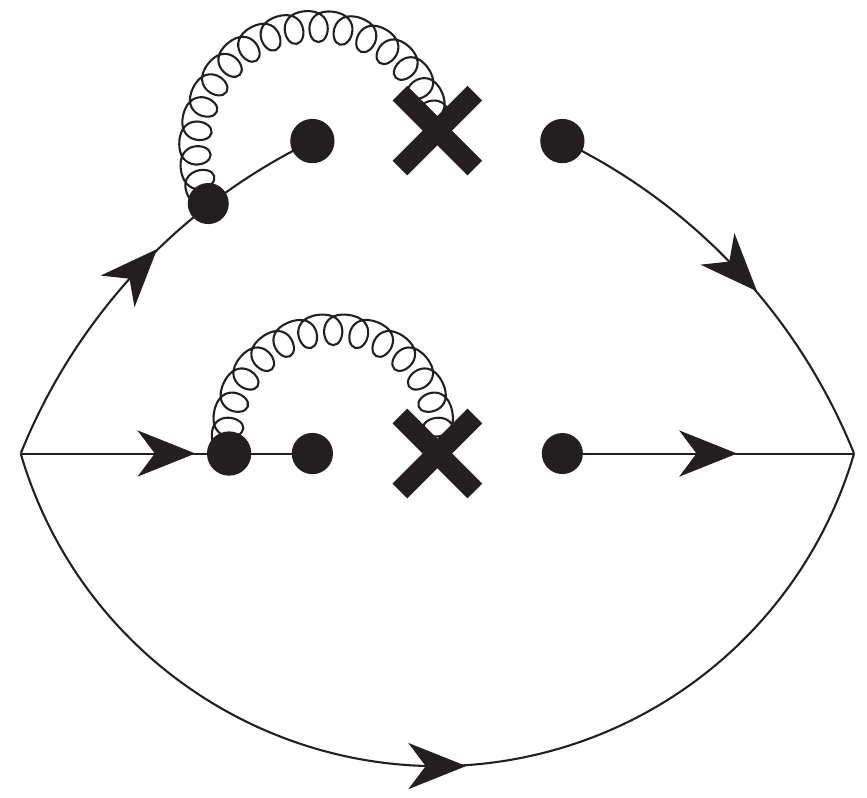}}
\\[2mm]
\scalebox{0.12}{\includegraphics{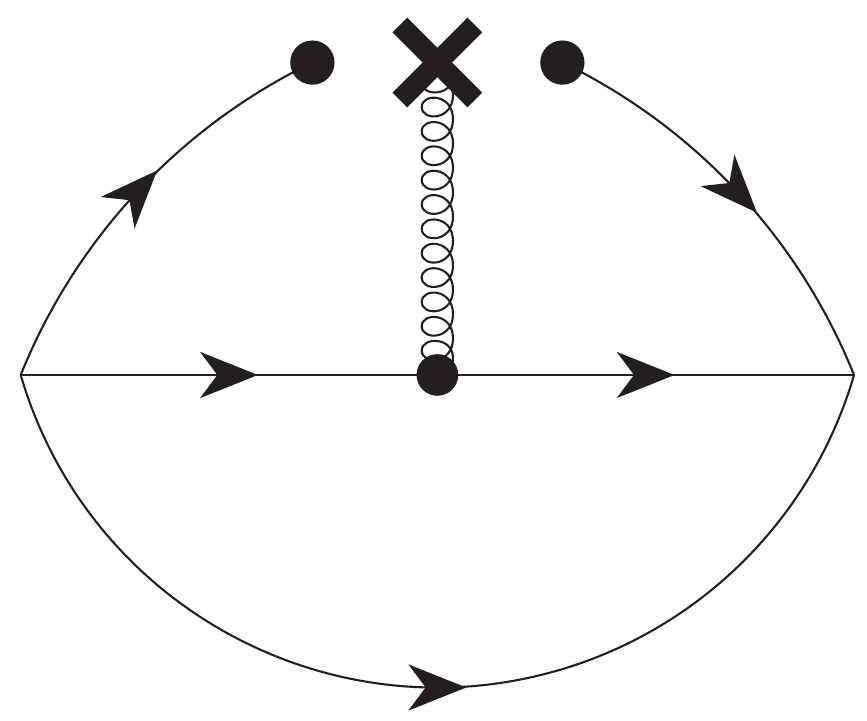}}~~
\scalebox{0.12}{\includegraphics{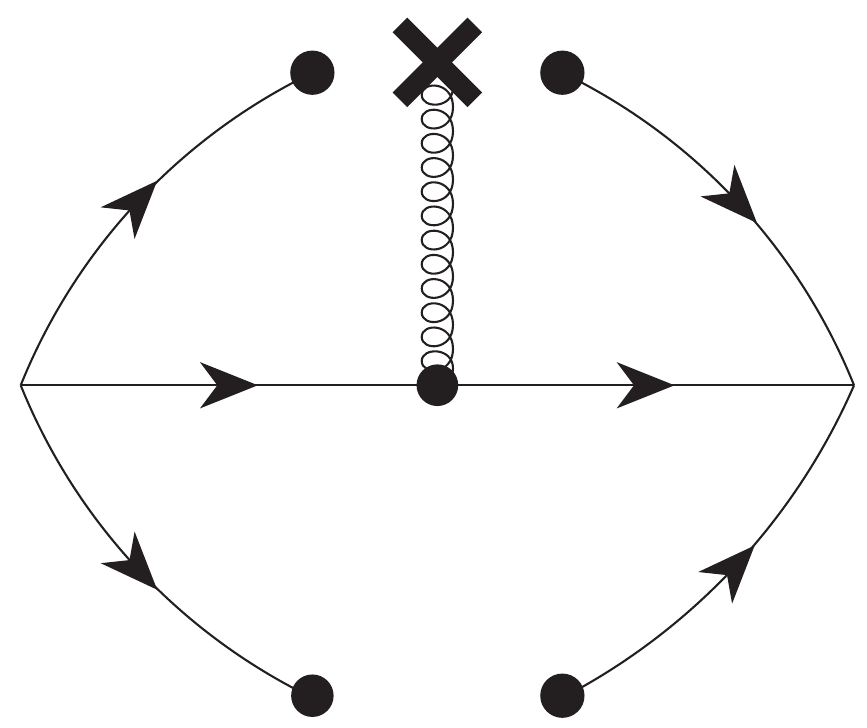}}~~
\scalebox{0.12}{\includegraphics{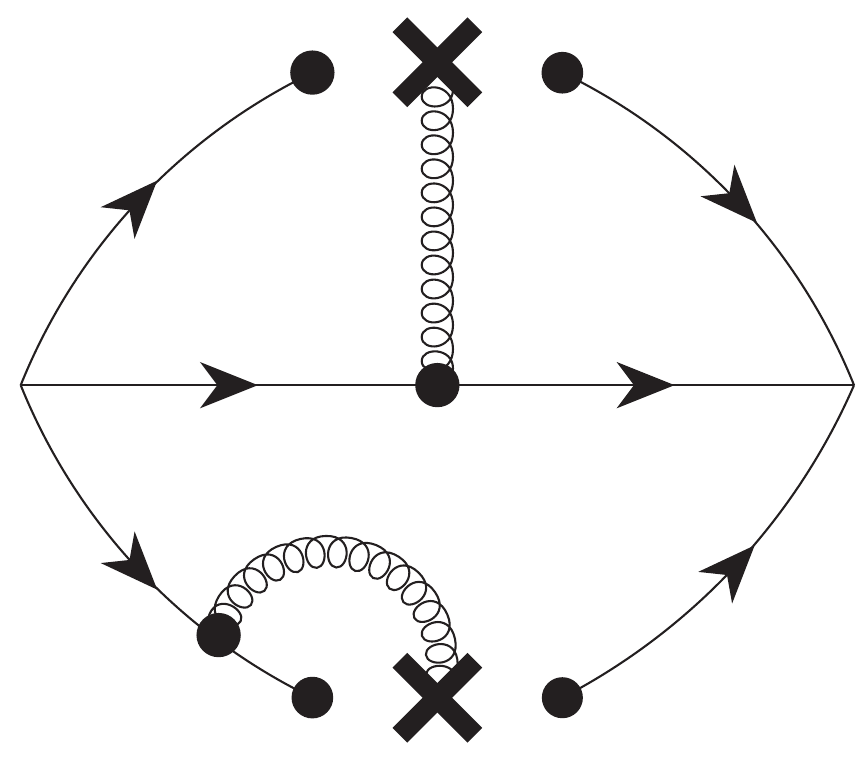}}
\\[2mm]
\scalebox{0.12}{\includegraphics{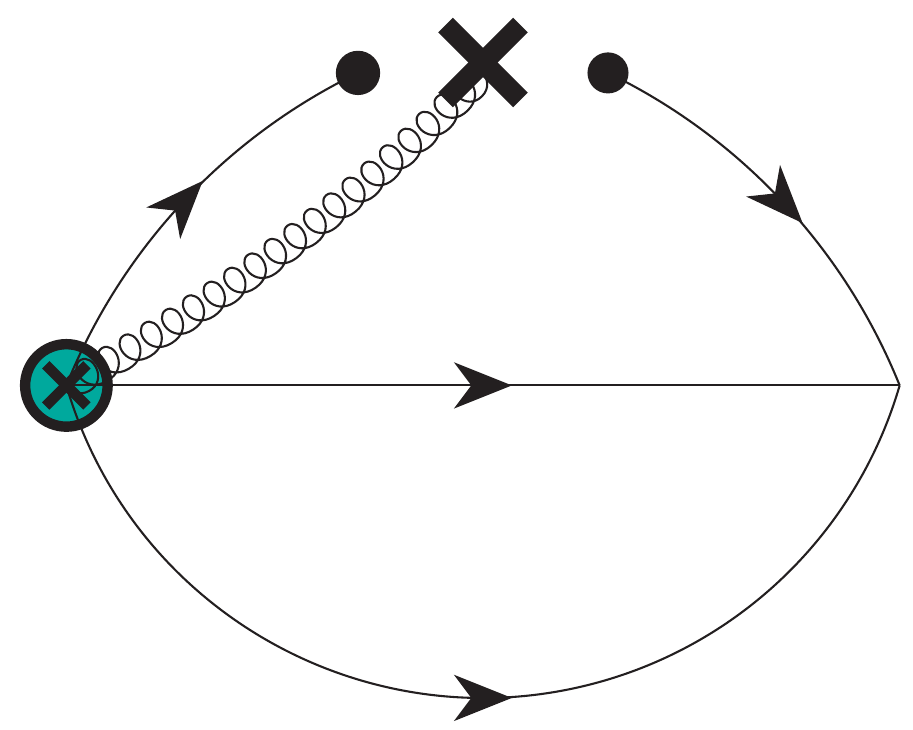}}~~
\scalebox{0.12}{\includegraphics{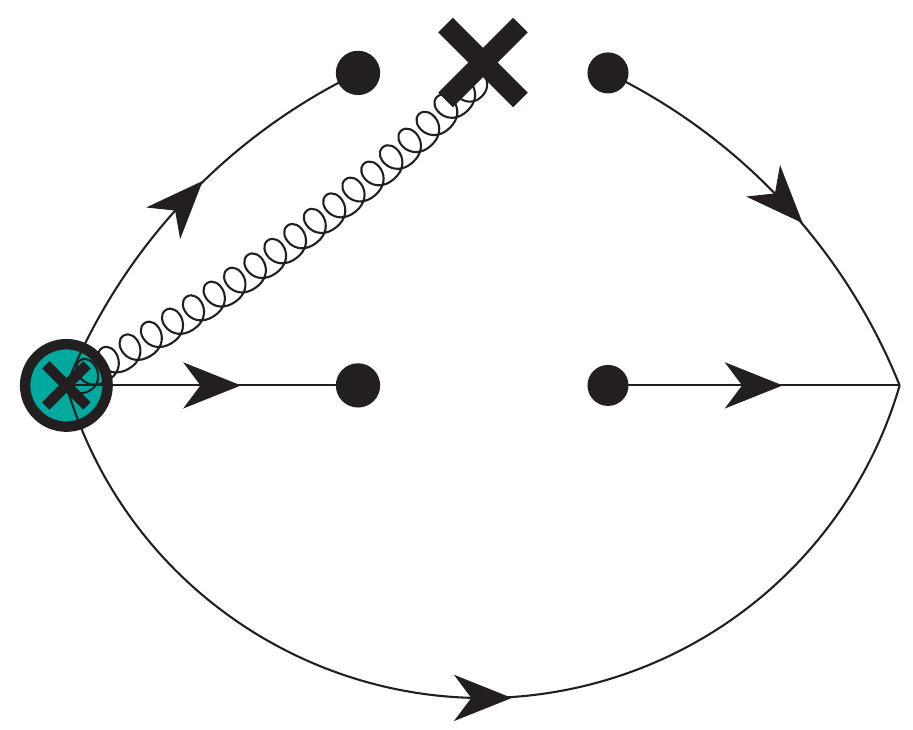}}~~
\scalebox{0.12}{\includegraphics{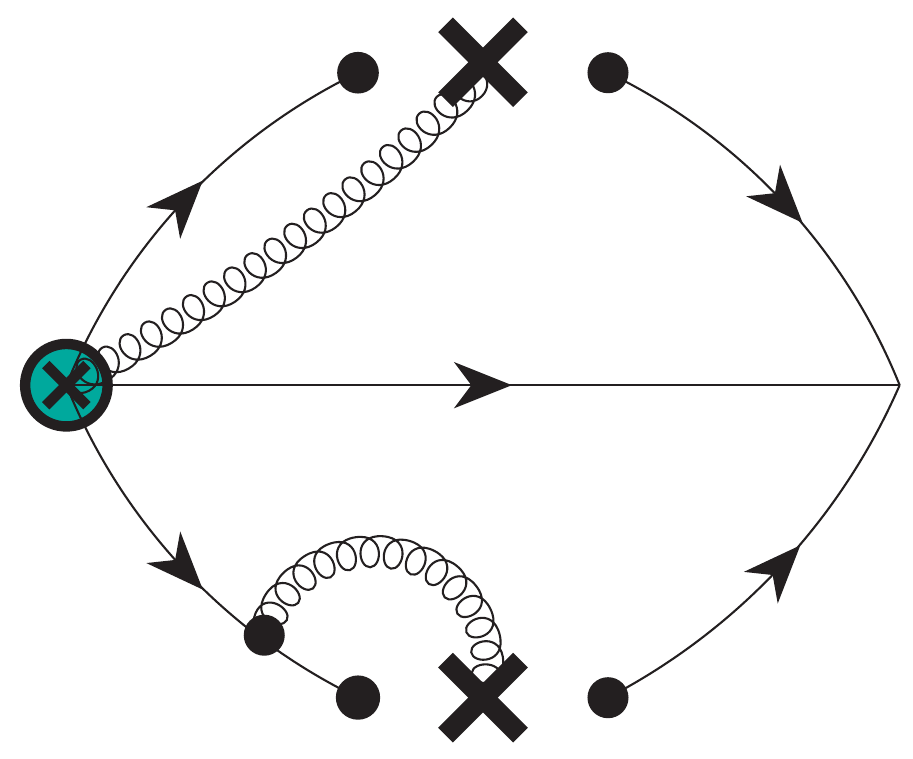}}
\\[2mm]
\scalebox{0.12}{\includegraphics{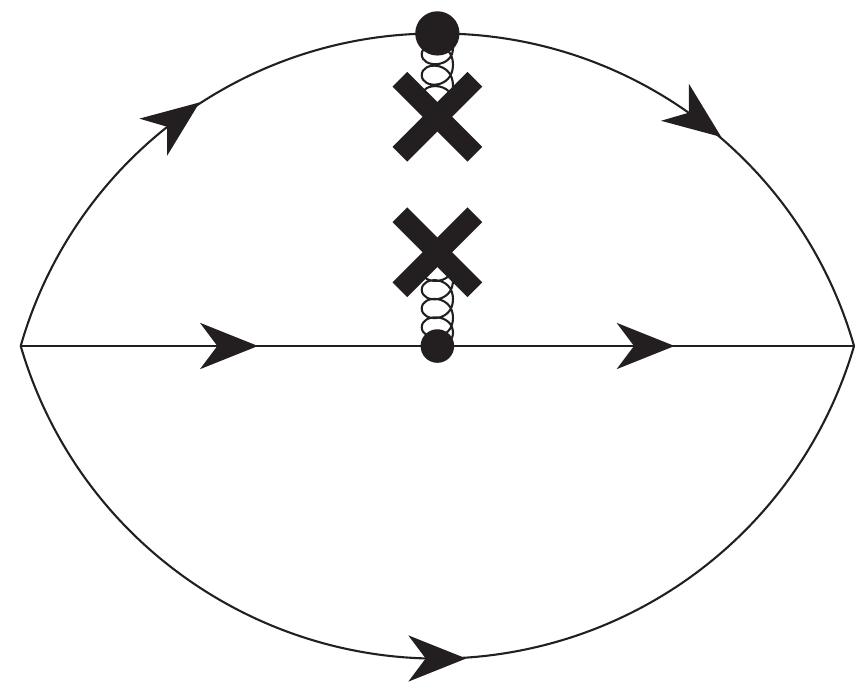}}~~
\scalebox{0.12}{\includegraphics{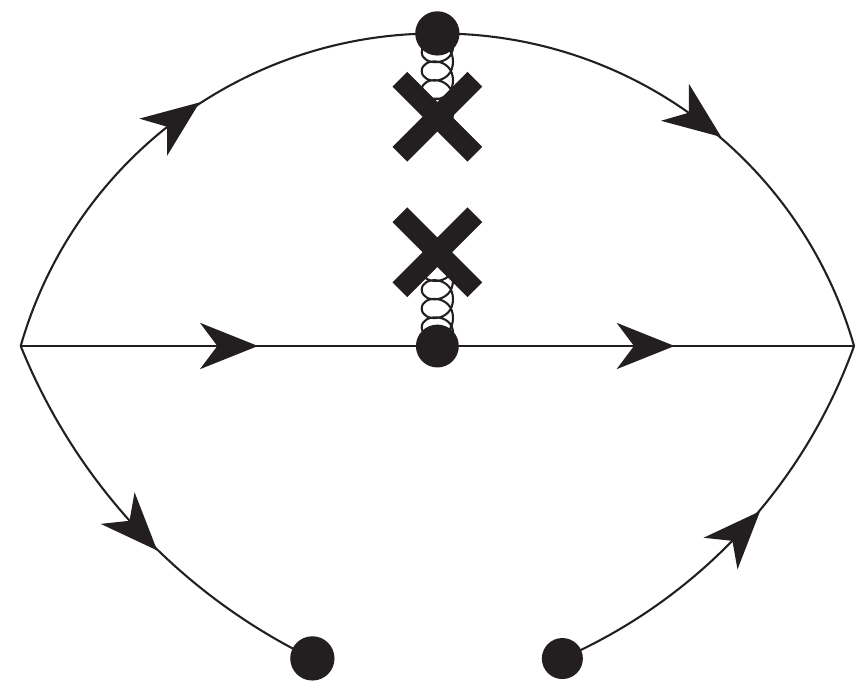}}~~
\scalebox{0.12}{\includegraphics{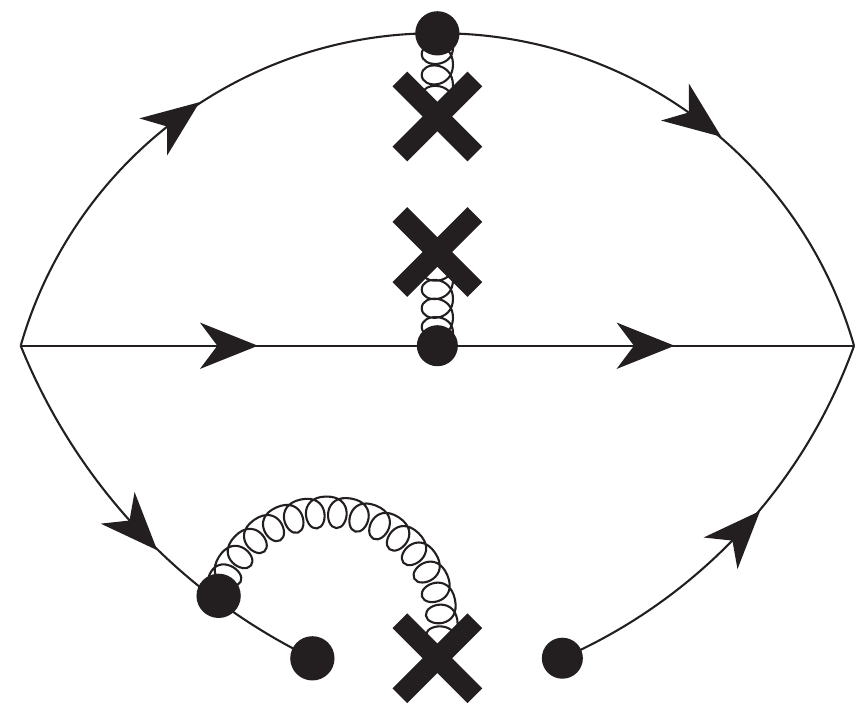}}
\\[2mm]
\scalebox{0.12}{\includegraphics{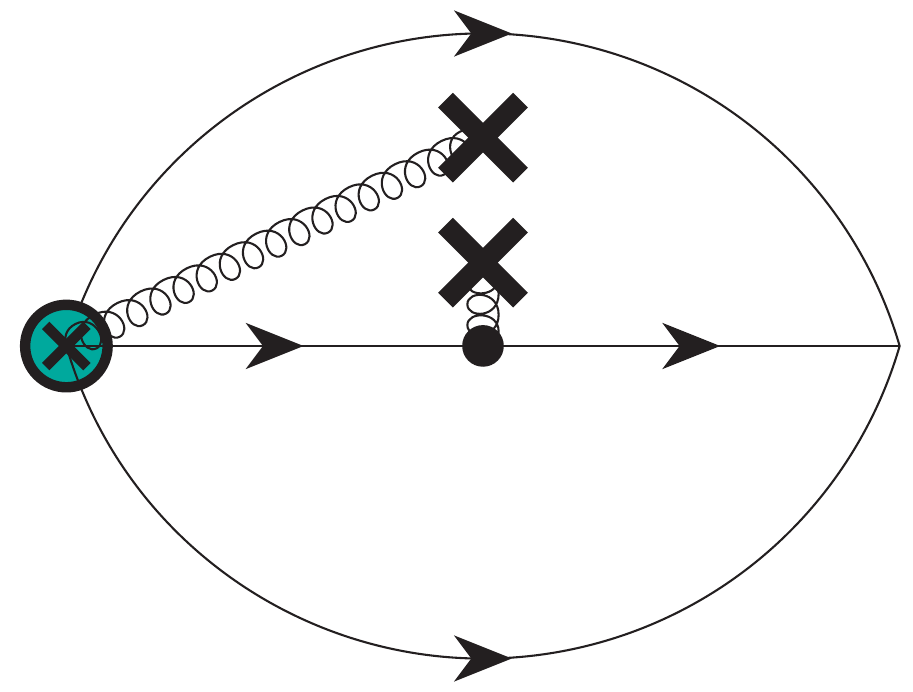}}~~
\scalebox{0.12}{\includegraphics{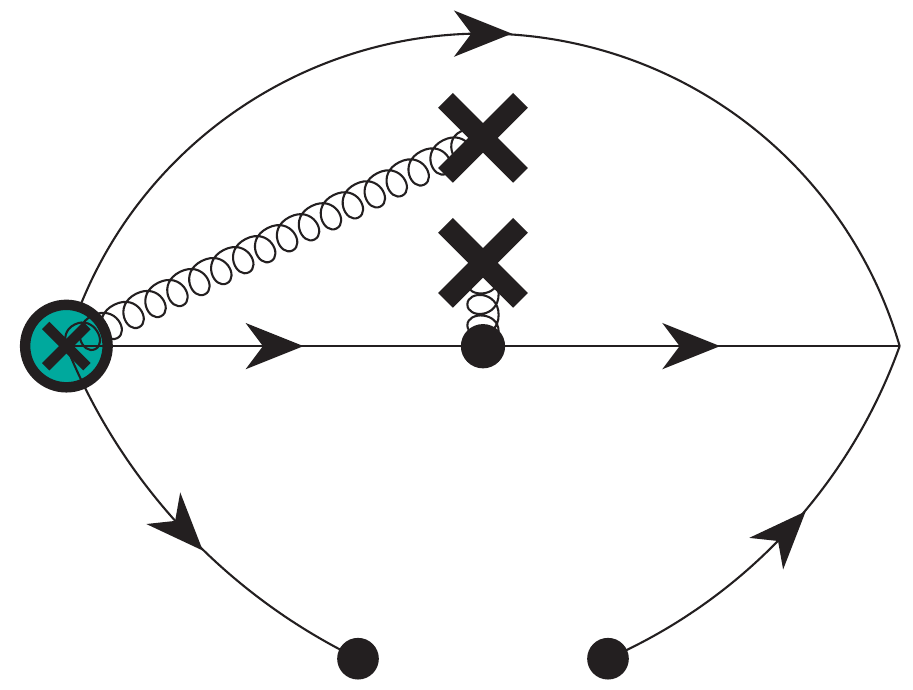}}~~
\scalebox{0.12}{\includegraphics{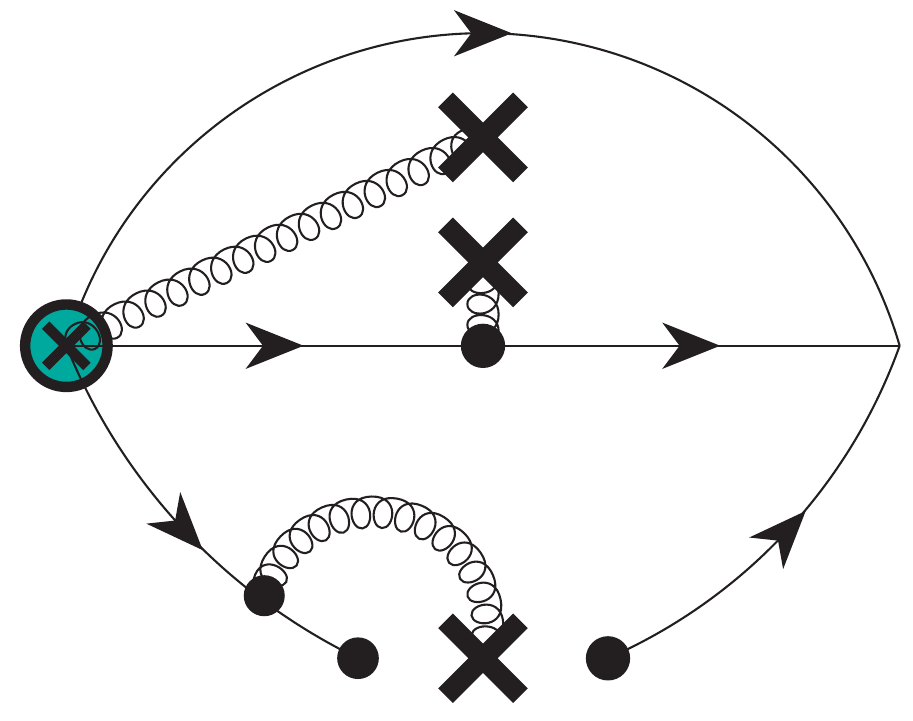}}
\caption{Feynman diagrams in the present study. The cross-circle (green) vertices represent the quark-gluon couplings due to the covariant derivative $D_\mu = \partial_\mu + i g_s A_\mu$.}
\label{fig:feynman}
\end{center}
\end{figure}

At the quark-gluon level, we use the method of the operator product expansion to calculate the correlation function in Eq.~(\ref{twoponit}), 
from which we extract the spectral densities $\rho^{\rm OPE}_1(s)$ and $\rho^{\rm OPE}_0(s)$ corresponding to Eq.~(\ref{phen1}) and Eq.~(\ref{phen0}) as 
\begin{eqnarray}
&& \nonumber\rho^{\rm OPE}_1(s)  
\\ &=& \label{OPE1}
{5 s^3 \over 36864 \pi^4}-{167 m_s^2 s^2 \over 40960 \pi^4}
\\  \nonumber&&
- \Big( {5 \langle g_s^2 GG \rangle \over 49152 \pi^4}
-{11 m_s \langle \bar s s \rangle \over 1024 \pi^2}\Big )s
\\ \nonumber &&
+ \Big ( - {263 m_s \langle g_s \bar s \sigma G s \rangle \over 73728 \pi^2}
+ { 155 m_s^2\langle g_s^2 GG \rangle  \over 196608 \pi^4 }\Big )
\\ \nonumber &&
-\Big ({89 \langle \bar s s \rangle \langle g_s \bar s \sigma G s \rangle \over 3072}
-{29 m_s\langle g_s^2 GG \rangle \langle \bar s s \rangle \over 147456 \pi^2}
\\ \nonumber &&
-{m_s^2\langle \bar s s \rangle^2\over128 } \Big )\delta(s)
+ \Big( -{361 \langle g_s \bar s \sigma G s \rangle^2 \over 36864} 
\\ \nonumber &&
-{\langle g_s^2 GG \rangle \langle \bar s s \rangle^2 \over 1152} + { m_s\langle g_s^2 GG \rangle \langle g_s \bar s \sigma G s \rangle \over 3072\pi^2}
\\ \nonumber &&
+{7 m_s^2 \langle \bar s s \rangle \langle g_s \bar s \sigma G s \rangle \over 192} \Big)\delta^\prime(s)\, ,
\\ 
&& \nonumber \rho^{\rm OPE}_0(s)  
\\ &=& \label{OPE0}
{23 m_s s^3 \over 32768 \pi^4}-{\langle \bar s s \rangle s^2 \over 96 \pi^2}
- \Big( {317 m_s \langle g_s^2 GG \rangle \over 589824 \pi^4}
\\  \nonumber&&
-{103 m_s^2 \langle \bar s s \rangle \over 1536 \pi^2}
+{37 \langle g_s \bar s \sigma G s \rangle \over 3072 \pi^2} \Big )s
\\ \nonumber &&
+ \Big (  {441 m_s^2 \langle g_s \bar s \sigma G s \rangle \over 8192 \pi^2}
+ { 25 \langle \bar s s \rangle \langle g_s^2 GG \rangle  \over 24576 \pi^2 }\Big )
\\ \nonumber &&
-\Big ({13 m_s^2\langle g_s^2 GG \rangle \langle \bar s s \rangle \over 4096 \pi^2}
-{13 \langle g_s^2 GG \rangle \langle g_s \bar s \sigma G s \rangle \over 294912\pi^2}
\\ \nonumber &&
+{5 m_s\langle \bar s s \rangle \langle g_s \bar s \sigma G s \rangle \over64 }\Big )\delta(s)
+ \Big( -{5 m_s \langle g_s \bar s \sigma G s \rangle^2 \over 2084} 
\\ \nonumber &&
-{3m_s\langle g_s^2 GG \rangle \langle \bar s s \rangle^2 \over 1024} + { 9m_s^2\langle g_s^2 GG \rangle \langle g_s \bar s \sigma G s \rangle \over 8192\pi^2} \Big)\delta^\prime(s)\, .
\end{eqnarray}
The corresponding Feynman diagrams are shown in Fig.~\ref{fig:feynman}. We ignore terms, in which all quark lines are cut and are only connected by one gluon propagator, because they are of higher order in $\alpha_s$. We preserve the $\rho^{\rm OPE}_1(s)$ up to dimension 10 (D=10) terms and $\rho^{\rm OPE}_0(s)$ up to dimension 11 (D=11) terms, including the perturbative term, the quark condensate $\langle \bar s s \rangle$, the gluon condensate $\langle g_s^2 GG \rangle$, the quark-gluon mixed condensate $\langle g_s \bar s \sigma G s \rangle$, and their combinations. 
Note that the four-quark condensate can be parametrized as $\langle \bar s s \bar s s\rangle$ $\sim$ $\kappa \langle \bar s s\rangle^2$, where $\kappa$ is the parameter that controls the violation of the vacuum factorization~\cite{Bertlmann:1984ih,Bertlmann:1987ty,Launer:1983ib,Narison:1992ru,Narison:1995jr,Narison:2009vy,Gubler:2015yna,Gubler:2016itj}. We will discuss the uncertainties induced by the violation of vacuum factorization in Section~\ref{sec:numerical}.   

By equating the spectral densities at the
hadron and quark-gluon levels, and
by performing the Borel transformation, we derive the sum rules as
\begin{eqnarray}
\Pi_\mp(s_0,M_B)&=& 2M_{\mp}f_{\mp}^2 e^{-M^{2}_{\mp}/M_B^2} \label{sumrule}
\\ \nonumber &=& \int^{s_0}_{s_<} (\sqrt{s} \rho^{\rm OPE}_1(s) \pm \rho^{\rm OPE}_0(s))e^{-s/M_B^2}ds .
\end{eqnarray}
Here the upper limit of the integral on the right hand side is set to $s_0$ under the assumption of the quark-hadron duality, 
namely that the phenomenological spectral density is equal to that of the OPE.   
The masses and coupling constants are obtained by the formulae
\begin{eqnarray}
\label{eq:mass} && M^2_{\mp}(s_0, M_B) 
\\ \nonumber&=& \frac{\int^{s_0}_{s_<} (\sqrt{s} \rho^{\rm OPE}_1(s) \pm \rho^{\rm OPE}_0(s)) s e^{-s/M_B^2} ds}{\int^{s_0}_{s_<} (\sqrt{s} \rho^{\rm OPE}_1(s) \pm \rho^{\rm OPE}_0(s)) e^{-s/M_B^2} ds} ,
\end{eqnarray}
and
\begin{eqnarray}
\label{eq:decay} &&f^2_{\mp}(s_0, M_B) 
\\\nonumber&=&\frac{\int^{s_0}_{s_<} (\sqrt{s} \rho^{\rm OPE}_1(s) \pm \rho^{\rm OPE}_0(s)) e^{-s/M_B^2} ds \times e^{M_{\mp}^2/M_B^2}}{2M_{\mp}} .
\end{eqnarray}

\section{Numerical analysis}
\label{sec:numerical}

In this section we present our numerical results.  
The used input 
parameters for condensates and masses determined at 
the renormalization scale 2~GeV are~\cite{pdg,Ovchinnikov:1988gk,Yang:1993bp,Ellis:1996xc,Ioffe:2002be,Jamin:2002ev,Gimenez:2005nt,Narison:2011xe,Narison:2018dcr}: 
\begin{eqnarray}
\langle\bar qq \rangle &=& -(0.240 \pm 0.010)^3 \mbox{ GeV}^3 \, ,\label{eq:condensate}
\\ \nonumber \langle\bar ss \rangle &=& (0.8\pm 0.1)\times \langle\bar qq \rangle \, ,
\\ \nonumber \langle g_s\bar s\sigma G s\rangle &=&  (0.8 \pm 0.2)\times\langle\bar ss\rangle \, ,
\\ \nonumber \langle \alpha_s GG\rangle &=& (6.35 \pm 0.35) \times 10^{-2} \mbox{ GeV}^4 \, ,
\\ \nonumber m_s &=& 93 ^{+9}_{-3} \mbox{ MeV} \, .
\end{eqnarray}

Let us first discuss the $J^P = 3/2^-$ state.  
From Eq.~(\ref{eq:mass}), it is understood that the mass depends on two free parameters: the Borel mass $M_B$ and the threshold value $s_0$. To find their proper working regions, we use three criteria: a) sufficiently good convergence of the OPE, 
b) sufficiently large pole contribution, and c) sufficiently weak mass dependence on these two parameters. 

\begin{figure}[hbt]
\begin{center}
\includegraphics[width=0.47\textwidth]{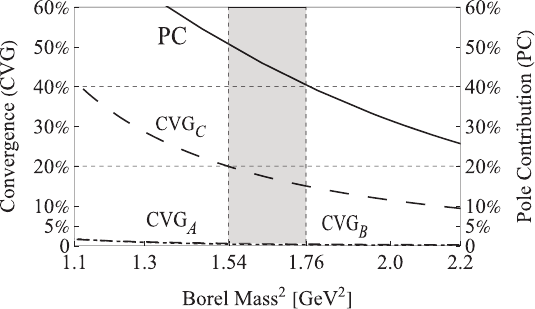}
\caption{CVG$_{A/B/C}$ and PC as functions of the Borel mass $M_B$ when $s_0$ is set to 6.0~GeV$^2$. CVG$_A$ (short-dashed) and CVG$_B$ (medium-dashed) are almost overlaid.}
\label{fig:cvgpole}
\end{center}
\end{figure}

To ensure the OPE convergence, we require that the higher dimensional contributions are sufficiently small, specifically
\begin{eqnarray}
\mbox{CVG}_A &\equiv& \left|\frac{ \Pi_-^{{\rm D=11+10+9+8}}(\infty, M_B^2) }{ \Pi_-(\infty, M_B^2) }\right| \leq 5\% \, ,
\\ 
\mbox{CVG}_B &\equiv& \left|\frac{ \Pi_-^{{\rm D=7+6}}(\infty, M_B^2) }{ \Pi_-(\infty, M_B^2) }\right| \leq 10\% \, ,
\\ 
\mbox{CVG}_C &\equiv& \left|\frac{ \Pi_-^{{\rm D=5+4}}(\infty, M_B^2) }{ \Pi_-(\infty, M_B^2) }\right| \leq 20\% \, ,
\label{eq:CVG_C}
\end{eqnarray}
where the denominator is the full OPE while the numerators are the contributions from the terms of dimensions as indicated 
by the upper labels.  
As shown in Fig.~\ref{fig:cvgpole} by the three dashed curves, we find that the Borel mass $M_B^2$ must be larger 
than 1.54 GeV$^2$, as determined by the $\mathrm{CVG}_C$ criterion of Eq.\,(\ref{eq:CVG_C}). 
Note that the other criteria are automatically satisfied in this case. 
For a sufficient pole contribution we require
\begin{equation}
\mbox{PC} \equiv \left|\frac{ \Pi_-(s_0, M_B^2) }{ \Pi_-(\infty, M_B^2) }\right| \geq 40\% \, .
\label{eq:pole_contribution_cond}
\end{equation}
As shown in Fig.~\ref{fig:cvgpole} by the solid curve, we find that the Borel mass $M_B^2$ must be less than 1.76 GeV$^2$ when $s_0 = 6.0$~GeV$^2$ for Eq.\,(\ref{eq:pole_contribution_cond}) to be satisfied. In the analysis of the above two criteria, we noticed that $s_0$ has a minimum value $s^{\rm min}_0 = 5.3$~GeV$^2$, and we have chosen $s_0$ slightly larger than it. Altogether the Borel window is determined to be $1.54$~GeV$^2 \leq M_B^2 \leq 1.76$~GeV$^2$ when $s_0 = 6.0$~GeV$^2$, and the working region of $s_0$ is determined to be $5.3$~GeV$^2 \leq s_0 \leq 6.7$~GeV$^2$.

\begin{figure*}[hbtp]
\begin{center}
\subfigure[]{\includegraphics[width=0.35\textwidth]{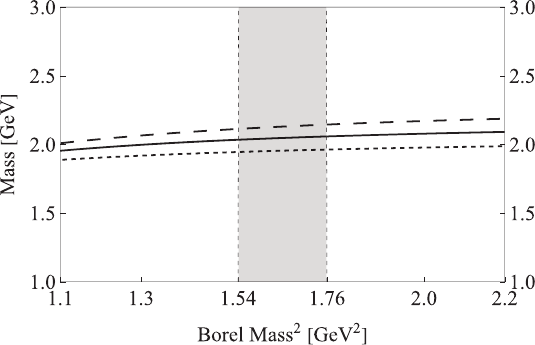}}
~~~~~~~~~~
\subfigure[]{\includegraphics[width=0.35\textwidth]{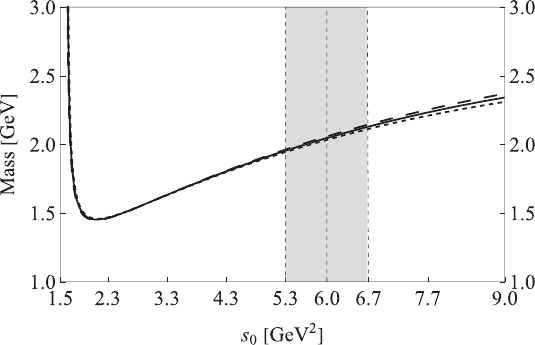}}
\caption{The mass $M_{3/2^-}$ as a function of the Borel mass $M_B^2$ and the threshold value $s_0$ extracted from the current $J_{\mu}$ in Eq~(\ref{def:current2}). In the left panel, the short-dashed/solid/long-dashed curves are obtained by setting $s_0 = 5.3/6.0/6.7$~GeV$^2$, respectively. In the right panel, the short-dashed/solid/long-dashed curves are obtained by setting $M_B^2 = 1.54/1.65/1.76$~GeV$^2$, respectively.}
\label{fig:mass}
\end{center}
\end{figure*}

We can now study the mass of the $3/2^-$  state as a function of the Borel mass $M_B^2$ and the threshold value $s_0$ as shown in Fig.~\ref{fig:mass}. From the left panel, we observe that the mass is almost independent of $M_B^2$ in the region $1.54$~GeV$^2 \leq M_B^2 \leq 1.76$~GeV$^2$. From the right panel, the mass dependence on $s_0$ is acceptable in the region $5.3$~GeV$^2 \leq s_0 \leq 6.7$~GeV$^2$. Note that the mass has a stability point at around $s_0 \sim$ 2.1 GeV$^2$ as shown in the right panel of Fig.~\ref{fig:mass}. However, the Borel window lies above this point; it exists only if $s_0 \geq s_0^{\rm min} = 5.3$~GeV$^2$. So we choose $s_0$ sightly large than this value.  The mass and coupling constant are calculated to be
\begin{eqnarray}
M_{3/2^-} &=& 2.05^{+0.09}_{-0.10}{\rm~GeV} \, ,
\label{eq:mass3-}
\\ \nonumber
f_{3/2^-} &=& 0.037^{+0.007}_{-0.007}{\rm~GeV^3} \, .
\label{eq:decay3-}
\end{eqnarray}
The uncertainties come from $s_0$, $M_B$ and the condensate list in Eq.~(\ref{eq:condensate}). We also consider the violation of the four-quark factorization where $\kappa$ is varied from 1 to 7~\cite{Gubler:2015yna,Gubler:2016itj}.  We have found that the uncertainties caused by it are very small and negligible. This is because the four-quark condensates appear only in the high dimension terms (D$\geq$8) in the OPE.

\begin{figure*}[hbtp]
\begin{center}
\subfigure[]{\includegraphics[width=0.35\textwidth]{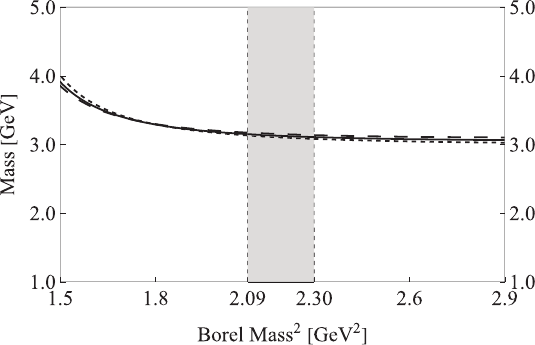}}
~~~~~~~~~~
\subfigure[]{\includegraphics[width=0.35\textwidth]{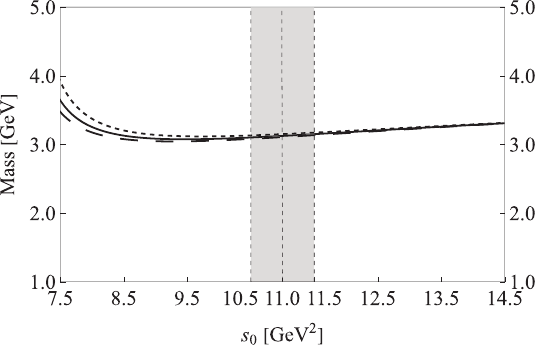}}
\caption{The mass $M_{3/2^+}$ as a function of the Borel mass $M_B^2$ and the threshold value $s_0$ extracted from the current $J_{\mu}$ in Eq~(\ref{def:current2}). In the left panel, the short-dashed/solid/long-dashed curves are obtained by setting $s_0 = 10.5/11.0/11.5$~GeV$^2$, respectively. In the right panel, the short-dashed/solid/long-dashed curves are obtained by setting $M_B^2 = 2.09/2.20/2.30$~GeV$^2$, respectively.}
\label{fig:mass3}
\end{center}
\end{figure*}

For the $J^P = 3/2^+$ state, we determine that its working regions are $2.09$~GeV$^2 \leq M_B^2 \leq 2.30$~GeV$^2$ and $10.5$~GeV$^2 \leq s_0 \leq 11.5$~GeV$^2$ as shown in Fig.~\ref{fig:mass3}, and calculate its mass and coupling constant as  
\begin{eqnarray}
M_{3/2^+} &=& 3.13^{+0.27}_{-0.18}{\rm~GeV} \, ,
\label{eq:mass3+}
\\ \nonumber
f_{3/2^+} &=& 0.074^{+0.015}_{-0.009}{\rm~GeV^3} \, .
\label{eq:decay3+}
\end{eqnarray}   
The mass of the positive parity state is estimated to be about 1 GeV larger than that of the negative parity state with larger uncertainties (which also holds for the coupling constant). This implies that the current with a derivative couples to and works for the negative parity state better than for the positive parity state. 

\begin{figure*}[hbtp]
\begin{center}
\subfigure[]{\includegraphics[width=0.35\textwidth]{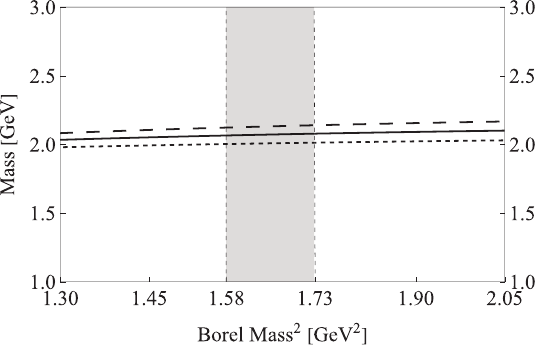}}
~~~~~~~~~~
\subfigure[]{\includegraphics[width=0.35\textwidth]{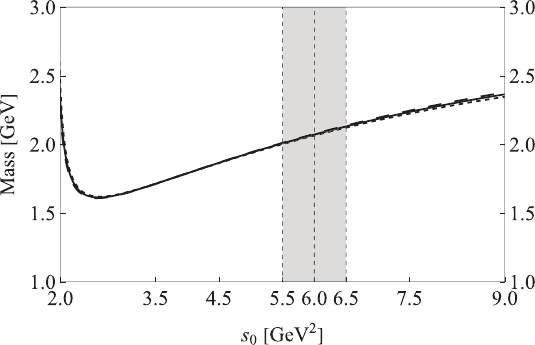}}
\caption{The mass $M_{1/2^-}$ as a function of the Borel mass $M_B^2$ and the threshold value $s_0$ extracted from the current $J$ in Eq~(\ref{def:current1}). In the left panel, the short-dashed/solid/long-dashed curves are obtained by setting $s_0 = 5.5/6.0/6.5$~GeV$^2$, respectively. In the right panel, the short-dashed/solid/long-dashed curves are obtained by setting $M_B^2 = 1.58/1.65/1.73$~GeV$^2$, respectively.}
\label{fig:mass1}
\end{center}
\end{figure*}

\begin{figure*}[hbtp]
\begin{center}
\subfigure[]{\includegraphics[width=0.4\textwidth]{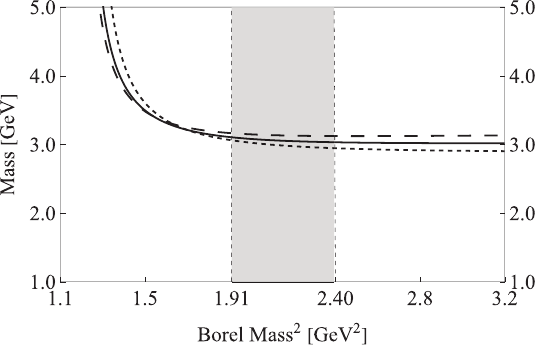}}
~~~~~~~~~~
\subfigure[]{\includegraphics[width=0.4\textwidth]{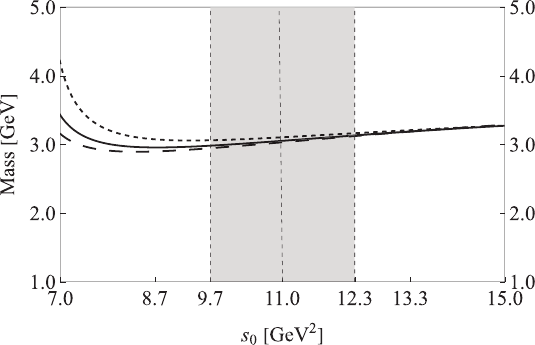}}
\caption{The mass $M_{1/2^+}$ as a function the Borel mass $M_B^2$ and the threshold value $s_0$ extracted from the current $J$ in Eq~(\ref{def:current1}). In the left panel, the short-dashed/solid/long-dashed curves are obtained by setting $s_0 = 9.7/11.0/12.3$~GeV$^2$, respectively. In the right panel, the short-dashed/solid/long-dashed curves are obtained by setting $M_B^2 = 1.91/2.16/2.40$~GeV$^2$, respectively.}
\label{fig:mass2}
\end{center}
\end{figure*}

We next perform the same numerical analysis using the current $J$ in Eq.~(\ref{def:current1}) to investigate the $J^P=1/2^\pm$ states. For the $J^P = 1/2^-$ state, the working regions are determined to be $1.58$~GeV$^2 \leq M_B^2 \leq 1.73$~GeV$^2$ and $5.5$~GeV$^2 \leq s_0 \leq 6.5$~GeV$^2$. We show its mass dependence on these two parameters in Fig.~\ref{fig:mass1}. The mass and the coupling constant are calculated to be
\begin{eqnarray}
M_{1/2^-} &=& 2.07^{+0.07}_{-0.07}{\rm~GeV} \, ,
\label{eq:mass1-}
\\ \nonumber
f_{1/2^-} &=& 0.079^{+0.011}_{-0.011}{\rm~GeV^3} \, .
\label{eq:decay1-}
\end{eqnarray}
For the $J^P = 1/2^+$ state, we show its working regions in Fig.~\ref{fig:mass2}, where we extract its mass and coupling constant as
\begin{eqnarray}
M_{1/2^+} &=& 3.05^{+0.21}_{-0.15}{\rm~GeV} \, ,
\label{eq:mass1+}
\\ \nonumber
f_{1/2^+} &=& 0.168^{+0.045}_{-0.040}{\rm~GeV^3} \, .
\label{eq:decay1+}
\end{eqnarray}
Once again we observe that the current couples better to the negative parity state.

The above results are summarized in Table~\ref{tab:results}. 
By using the three-quark currents with a derivative supplemented by spin and parity projection, $J$ and $J_\mu$ in Eqs.~ (\ref{def:current1}) and (\ref{def:current2}) respectively, we have obtained the masses of $J^P = 1/2^-$ and $3/2^-$ states in a mass region around 2~GeV, while the positive parity states appear in a mass region around 3~GeV.  

In all, our results therefore suggest that the recently observed $\Omega(2012)$ baryon is likely to have negative parity.  To further determine its spin, we need to study its decay properties. 
Naively, because of the centrifugal barrier existing only for the higher spin state, we expect that the $J^P = 1/2^-$ ($3/2^-$) state has a wider (narrower) decay width, and therefore, the observed state with a width around 6 MeV is likely to be a $J^P = 3/2^-$ state.  

To do a complete analyses, we have also performed an investigation by using a current without derivative~\cite{Chen:2012ex}: 
\begin{eqnarray}
J^\prime_{\mu} = -\sqrt{3} \epsilon^{abc} s^T_aC\gamma_\mu s_b s_c \, , 
\label{current3}
\end{eqnarray}
whose spin parity is $3/2^+$. Once again, we have performed sum rule analysis by setting the proper criteria for convergence of OPE and pole contribution, with the parity projection properly done as before. As summarised in Table~\ref{tab:results}, the extracted masses are  
\begin{eqnarray}
M^\prime_{3/2^+} &=& 1.59^{+0.10}_{-0.12}{\rm~GeV} \, ,
\\
M^\prime_{3/2^-} &=& 3.15^{+0.16}_{-0.17}{\rm~GeV} \, .
\end{eqnarray}
The mass of $3/2^+$ is in good agreement with the $\Omega(1672)$. 
Once again, the uncertainty is larger for higher energy state of negative parity.
The sum rule for  higher energy states in the present method could not be efficiently performed.  
In this respect, an alternative form of sum rule exclusively for the negative parity state with the mass of the positive parity state treated as an input could be another option~\cite{Aliev:2018syi}.


\begin{table*}[hpt]
\begin{center}
\renewcommand{\arraystretch}{1.25}
\caption{Masses and coupling constants extracted from the currents $J$ in Eq.~(\ref{def:current1}), $J_{\mu}$ in Eq.~(\ref{def:current2}), and $J^\prime_\mu$ in Eq.(~\ref{current3}).}
\begin{tabular}{c|c|c|c|c|c|c|c}
\hline\hline
~\multirow{2}{*}{Current}~& ~~\multirow{2}{*}{state}~~ & ~\multirow{2}{*}{~$s_0^{\rm min}~[{\rm GeV}^2]$~}~ & \multicolumn{2}{c|}{Working Regions} & ~\multirow{2}{*}{Pole~[\%]}~ & ~\multirow{2}{*}{~Mass~[GeV]~}~&~\multirow{2}{*}{~Couple constant~[GeV$^3$]~}~
\\ \cline{4-5}
&&&~~$M_B^2~[{\rm GeV}^2]$~~&~$s_0~[{\rm GeV}^2]$~~&&&
\\ \hline\hline
$J$&$|\Omega;1/2^+\rangle$&9.7&$1.91$-$2.40$&$11.0$&$40$-$55$&$3.05^{+0.21}_{-0.15}$&$0.168^{+0.045}_{-0.040}$
\\
&$|\Omega;1/2^-\rangle$ &5.5&$1.58$-$1.73$&$6.0$&$40$-$47$&$2.07^{+0.07}_{-0.07}$&$0.079^{+0.011}_{-0.011}$
\\
$J_\mu$&$|\Omega;3/2^+\rangle$&10.5&$2.09$-$2.30$&$11.0$&$40$-$46$&$3.13^{+0.27}_{-0.18}$&$0.074^{+0.015}_{-0.009}$
\\
&$|\Omega;3/2^-\rangle$ 
&5.3&$1.54$-$1.76$&$6.0$&$40$-$51$&$2.05^{+0.09}_{-0.10}$&$0.037^{+0.007}_{-0.007}$
\\
$J^\prime_\mu$&$|\Omega^\prime;3/2^+\rangle$&3.3&$1.48$-$1.77$&$4.0$&$40$-$52$&$1.59^{+0.10}_{-0.12}$&$0.033^{+0.006}_{-0.006}$
\\
&$|\Omega^\prime;3/2^-\rangle$ &11.5&$3.30$-$3.93$&$13.0$&$40$-$51$&$3.15^{+0.16}_{-0.17}$&$0.092^{+0.018}_{-0.018}$
\\ \hline\hline
\end{tabular}
\label{tab:results}
\end{center}
\end{table*}

\section{summary}
\label{sec:summary}


In this paper, we studied the recently observed $\Omega$ baryon $\Omega(2012)$ making use of QCD sum rules. We constructed the $P$-wave $\Omega$ baryon currents with a covariant derivative, whose spins are 1/2 and 3/2 by performing the proper spin projections. We then analyzed the parity-projected QCD sum rules to separate the contribution of the positive parity and negative parity states. Thus, we systematically studied in total four states with spin-parity $1/2^\pm$ and $3/2^\pm$, and applied the QCD sum rule method to calculate their masses and coupling constants. 
The results are summarised in Table~\ref{tab:results}. We determined the mass of the $J^P = 1/2^-$ state as  
\begin{eqnarray}
M_{1/2^-} &=& 2.07^{+0.07}_{-0.07}{\rm~GeV} \, ,
\label{eq:result_m_12}
\end{eqnarray}
and that of $J^P = 3/2^-$ as 
\begin{eqnarray}
M_{3/2^-} &=& 2.05^{+0.09}_{-0.10}{\rm~GeV} \, .
\label{eq:result_m_32}
\end{eqnarray}
As both masses are consistent with the $\Omega(2012)$, it is likely that the $\Omega(2012)$ is a $P$-wave excited $\Omega$ baryon with three strange quarks. However, due to the closeness of Eqs.~(\ref{eq:result_m_12}) and (\ref{eq:result_m_32}), we cannot determine its spin quantum number in the present analysis. 

We have so far only studied the mass and coupling constant of the $\Omega(2012)$ baryon. Based on these results, we plan to investigate its decay properties in the near future, which are also important to understand its internal structure. If its spin parity is $J^P = 3/2^-$, it may more likely decay via S-wave to the final state $\bar K \Xi(1530)$, even though such a decay will be suppressed due to a small phase space factor. 
It can also decay to the final $\bar K \Xi$ state, but only as a $D$-wave final state, so that the total decay width should be small. On the other hand, if its spin parity is $J^P = 1/2^-$, it will be easier to decay to $\bar K \Xi$ via $S$-wave, for which there is no phase space suppression, while the decay to $\bar K \Xi(1530)$ proceeds via $D$-wave, 
so that the total decay width should become much larger. 
All these naive expectations need to be confirmed by an actual QCD sum rule analysis. 
In addition, QCD sum rule analysis using a five-quark current corresponding to the molecular state $\bar K \Xi(1530)^{\ast}$ will also have to be carried out in the future.

\section*{Acknowledgments}
N.S. is supported by the China Scholarship Council under Grant No.~202306090272. 
H.X.C. is supported by
the National Natural Science Foundation of China under Grant No.~12075019,
the Jiangsu Provincial Double-Innovation Program under Grant No.~JSSCRC2021488,
and
the Fundamental Research Funds for the Central Universities. 
P.G. is supported by KAKENHI under Contract No. JP22H00122. 
A.H. is supported in part by the Grants-in-Aid for Scientific Research [Grant No. 21H04478(A), 24K07050(C)].

\bibliographystyle{elsarticle-num}
\bibliography{ref}

\end{document}